\documentclass[sigconf]{acmart}
\usepackage{adjustbox}
\usepackage{hyperref} \hypersetup{colorlinks=true, urlcolor=blue, linkcolor=red}

\AtBeginDocument{%
  \providecommand\BibTeX{{%
    \normalfont B\kern-0.5em{\scshape i\kern-0.25em b}\kern-0.8em\TeX}}}

\setcopyright{acmcopyright}
\copyrightyear{2020}
\acmYear{2020}
\acmDOI{10.1145/1122445.1122456}

\acmConference[EDM 2020]{Educational Data Mining}{July 10--13, 2020}{Ifrane, Morocco}
\acmBooktitle{EDM 2020: Educational Data Mining Conference,
  July 10--13, 2020, Ifrane, Morocco}
\acmPrice{15.00}
\acmISBN{978-1-4503-9999-9/18/06}

\newcommand{\tsc}[1]{\textsuperscript{#1}} 
\author{Yunkai Xiao [yxiao28],\tsc{1} Gabriel Zingle [gzingle],\tsc{1} Qinjin Jia [qjia3],\tsc{1} Harsh R. Shah [hshah3],\tsc{1}}
\author{Yi Zhang [20171184],\tsc{2} Tianyi Li [tianyi.li],\tsc{3} Mohsin Karovaliya [mrkarova],\tsc{1}}
\author{Weixiang Zhao [20172986],\tsc{2} Yang Song [songy],\tsc{4}  Jie Ji [20172986],\tsc{5}} \author{Ashwin Balasubramaniam [abalasu4],\tsc{1} Harshit Patel [hpatel24],\tsc{1}}
\author{Priyankha Bhalasubbramanian [pbhalas],\tsc{1}
Vikram Patel [vpatel22],\tsc{1} and Edward Gehringer [efg]\tsc{1}}

\affiliation{
  \institution{\vskip .2cm}
  \institution{1. North Carolina State University, Raleigh, North Carolina 27695, USA [@ncsu.edu]}
  \institution{2. Northeastern University, Shenyang, Liaoning 110819, China [@stu.neu.edu.cn]}
  \institution{3. Shanghai Jiao Tong University, Shanghai 201101, China [@sjtu.edu.cn]}
  \institution{4. University of North Carolina at Wilmington, Wilmington, North Carolina 28407, USA [@uncw.edu]}
  \institution{5. Southern University of Science and Technology, Shenzhen, Guangdong 518055, China [@mail.sustech.edu.cn]}
}

\begin{document}

\title{Detecting Problem Statements in Peer Assessments \\
(extended version)}
\thanks{\textit{Note:} This is an extended version of the paper of the same name published at the \underline{\href{http://educationaldatamining.org/edm2020/}{Educational Data Mining 2020}} conference.}

\begin{abstract}
  Effective peer assessment requires students to be attentive to the deficiencies in the work they rate.  Thus, their reviews should identify problems.  But what ways are there to check that they do? We attempt to automate the process of deciding whether a review comment detects a problem.  We use over 18,000 review comments that were labeled by the reviewees as either detecting or not detecting a problem with the work. We deploy several traditional machine-learning models, as well as neural-network models using GloVe and BERT embeddings. We find that the best performer is the Hierarchical Attention Network classifier, followed by the Bidirectional Gated Recurrent Units (GRU) Attention and Capsule model with scores of 93.1\% and 90.5\% respectively. The best non-neural network model was the support vector machine with a score of 89.71\%. This is followed by the Stochastic Gradient Descent model and the Logistic Regression model with 89.70\% and 88.98\%.
\end{abstract}



\maketitle
\pagestyle{plain}

\section{Introduction}
Peer assessment---students giving feedback on each other's work---has been a common educational practice for at least 50 years \cite{topping1998peer, tai20195}  It provides students more copious and rapid feedback than an instructor would give, as well as reactions from a more authentic audience (the student's peers).  By concentrating on a limited number of works, peers can produce assessments with similar validity and reliability to those of instructors, whose time is spread more thinly over many students' submissions \cite{topping2009peer}.  Students who perform peer assessment show a substantial increase in performance \cite{li2019does}.  Moreover, studies uniformly report that students learn more by being reviewers than they learn from the reviews they receive \cite{lundstrom2009give, ccevik2015assessor, li2010assessor, van2017exploring}.

The need for peer assessment was felt more acutely after the rise of massive open online courses (MOOCs) in the early part of this decade. With students paying minimal, if any, fees, MOOCs are not able to hire staff to serve as assessors for all submitted work. Thus, MOOCs rely heavily on peer assessment \cite{kay2013moocs, piech2013tuned}.

In order to gain much from peer assessment, students must take the process seriously.  They must think carefully and metacognitively about the works they are reviewing. To foster an atmosphere where students assess conscientiously, the instructor must offer some training in reviewing---and follow up by assessing how well the students perform this task \cite{liu2014assessment}.  But instructor assessment of students' reviewing suffers from the same shortcomings as instructor assessment of students' submitted work: it consumes much instructor time, is likely to be rushed, and is mostly summative; that is it evaluates how well the students have done, but does not directly help them improve their reviewing.  Thus, considerable research has been done on other methods for assessing review quality \cite{Gehringer2014survey}.

Fundamentally, the quality of a review is related to whether it identifies ways for the author to improve the work.  Thus, it is important for the review to point out shortcomings or problems the reviewer perceives in the reviewed work.  This paper describes several approaches to automatically identifying whether review \textit{comments}, which are responses to individual rubric items, do point out (alleged) problems with the work.

\section{Related Work}
One of the earliest approaches to improving review quality was calibration.  A student is asked to review work that is also being reviewed by a member of the course staff.  The review rubric contains some quantitative elements like Likert-scale items or checkboxes.  The online system computes how close the student's score is to the expert's.  The student may then be asked to review and score other works.  If the system is oriented toward peer grading---where students assign grades to other students---it may use the calibration score to weight this student's review's contribution to the author's final grade \cite{chapman2000calibrated}.  Alternatively, the system may continue assigning other work to review until the student reviewer comes "close enough" to the expert assessment \cite{piech2013tuned}.  A major advantage of calibration is that it saves time for the instructor, who does not have to manually rate the students' reviews.  However, it does demand extra time of the student.  In Calibrated Peer Review \cite{chapman2000calibrated}, for example, each student assesses three calibration instruments and three works by fellow students, meaning that half the student's time is spent providing feedback that will never be seen by another human.  Calibration is still the subject of active research \cite{wang2019sspa}.

Another quantitative approach to rating reviews is a reputation system \cite{hamer2005method, cho2007scaffolded}.  It is based on a rubric with quantitative elements as well.  But instead of comparing a student-assigned score with the instructor's, it compares the student-assigned score with scores assigned by other students who reviewed the same work.  

Most reputation systems include (at least) two metrics: leniency and spread.  Leniency measures whether this student reviewer tends to assign higher or lower scores than his/her peers.  Spread measures the difference between scores that the same reviewer assigns to different work. In general, a reviewer who has average leniency and high spread is rated most competent by a reputation system.  An advantage of reputation systems is that they can compute a metric of review quality without instructor intervention.  A disadvantage is that reputation systems work entirely on numerical (summative) scores, and do not encourage reviewers to provide helpful comments to their reviewees.  It might be said that reputation systems reward those with insight, regardless of their effort.

An alternative to having the instructor assess reviews is to have students assess reviews.  Such an assessment is sometimes called a "back-review" or a rejoinder.  Student reviewers can be graded based on how helpful their authors found their reviews.  However, a moral hazard of this approach is retaliation: a student reviewer may worry that if (s)he doesn't rate the author's work highly, the author won't rate his/her review highly either.  A variety of mitigation strategies have been devised. Crowdgrader \cite{de2014crowdgrader}, for example, drops the lowest one-quarter and highest one-quarter of back-review scores when calculating a student's reviewing grade.  If retaliation is not an issue, then back-reviews can be an effective way of assessing reviews.  The back-review rubric can include prompts that elicit feedback on how helpful the review was to the author.

There has also been considerable work on automating various aspects of review assessment with natural language processing (NLP) and machine-learning (ML) algorithms.  Brun and Hag\`{e}ge \cite{brun2013suggestion} applied NLP techniques to identify suggestions in review text.  Zingle et al. \cite{zingle2019detecting} showed that ML approaches could outperform NLP approaches for suggestion detection.  Xiong et al. \cite{xiong2011automatically, xiong2012natural} tried to automatically detect feedback that was localized, i.e., that referred to a particular location in the text.  Xiao et al. \cite{xiao2018application} applied neural networks to labeling review comments that contained several different kinds of content, such as praise, mitigation, and localization. Nguyen et al. \cite{nguyen2016instant} used logistic regression to train a model that predicted whether a review comment contained a problem solution.  They provided this information to the reviewer before the review was submitted, in order to encourage the reviewer to suggest solutions for problems in the work.

Peer assessment has much in common with peer review, as used to vet scientific work for publication.  Hua et al. \cite{hua-etal-2019-argument} annotated 400 scientific reviews and used them to automatically detect arguments that reviewers made in analyzing the research that the papers described.  Similar techniques could be applied to peer assessments.  This would facilitate identification of reviews that made judgments about the quality of a work without offering concrete reasons.  Similarly, automated detection of problems, solutions, and sentiment are of interest to help vendors to learn from online reviews of products and services.  Negi \cite{negi2017suggestion} has employed rule-based classifiers, SVMs, and neural networks to detect suggestions in product reviews.

\section{Experimental Methodologies}
\subsection{Data}

The data used for our experiments comes from Expertiza \cite{gehringer2010expertiza}, a peer-assessment platform oriented toward reviewing work developed by collaborative teams.  Reviews are performed by individuals; each individual reviews the work of some number of teams, typically 2 to 4.  For each review, the reviewer fills out a rubric, which consists of several criteria. Table \ref{tab:rubric} shows several sample rubric criteria.  Most criteria ask for a numeric rating as well as textual feedback. It is the textual feedback that we use in this work.   

\begin{table*}[htb]
  \caption{Sample Rubric Criteria}
  \label{tab:rubric}
  \begin{tabular}{ccl}
    \toprule
    Rating&Rubric Item\\
    \midrule
     1-5 & How well does the code follow "good Ruby and Rails coding practices"?\\
     1-5 & Is the user interface intuitive and easy to use? If not, is it well described in the README file?\\
    Yes/No & Can the admin delete users (customers or agents) from the system?\\
  \bottomrule
\end{tabular}
\end{table*}

Our study is based on reviews of coding and documentation assignments from NCSU CSC 517, Object-Oriented Design and Development. A typical assignment consists of these tasks: Instructor posts an assignment for students; students form teams of 2 to 4 members, depending on the assignment. After the teams submit their work, every member from each team is given submissions from two other teams to review. After the reviews are done, the teams revise their work based on the review feedback, and resubmit it. Then a second round of review is performed, using a second, summative, rubric, which asks questions about how well the team has addressed the feedback given in the first review round, and about the quality of the final submission.

Our neural networks and traditional machine-learning approaches require labeled data.  It would take the research team considerable time to tag enough data, and the tedium of the task might affect the reliability of tagging.  Our approach was to engage the students in labeling.  We offered the students a small amount of extra credit if they would tag each review comment they received as either detecting a problem or not detecting a problem. Since these reviews are on works the students have submitted, the students are in an ideal position to know whether the review comment mentioned a problem with their work. The course staff and the research team spot-checked the labels that the students tagged their reviews with, in order to award credit fairly and avoid polluting the data with random tags.  If the tags were found to be inaccurate, all tags submitted by that student were removed from the dataset. Some sample data can be found in Table \ref{tab:data}. 

\begin{table*}[!htbp]
  \caption{Sample Data}
  \label{tab:data}
  \begin{tabular}{ccl}
    \toprule
    Problem Statement?&Review Comment\\
    \midrule
     0 & the interface is easy to use and it is well described in the Readme file.\\
     1 & The implementation can only log one type of user on.\\
    1 & The titles and order need to change. The content needs to be placed at proper places.\\
  \bottomrule
\end{tabular}
\end{table*}

Since the reviews were done on team projects, and tagging was done individually, two to four students had the opportunity to tag the same review comments.  If multiple students did tag the same comment, inter-rater reliability (IRR) could be calculated.  We used Krippendorff's $\alpha$ \cite{Krippendorff2013Content} as the metric for IRR.   

The advantage of using Krippendorff's $\alpha$ compared to another IRR metric such as Cohen's kappa is that Krippendorff's $\alpha$ is not impacted by missing ratings, which often show up in students' peer-review data, since extra credit does not incentivize all students to annotate all their review comments.

The data that we pulled from Expertiza contained multiple tags for the same peer review comment by different raters. We included the comments with only a single tag, and those with multiple tags that matched unanimously. By dropping observations with conflicting tags by separate taggers, we have raised the Kirppendorff's $\alpha$ associated with our dataset from 0.696 to 1. 

The dataset was then de-duplicated and balanced through down-sampling. Initially, we had 9,177 positive tags and 9,490 negative tags. Down-sampling reduced the false tag count to 9,177 to match the quantity of positively tagged observations. This resulted in a class-balanced dataset with a total of 18,354 observations.

The dataset was separated into training, validation, and testing sets in the ratio of 80:10:10. This split was used to find optimized hyperparameters with 5-fold crossvalidation for the classifiers we used to identify a peer-review comment that detected a problem. Unless the dataset is large, the combination of observations used in the training and test sets can have an impact on how well the classifier performs. We compensated for this by using 20-fold cross validation on our finalized classifiers with tuned hyperparameters and saving the resulting 20 scores for analysis. This allowed us to generate summary statistics of each classifier's performance in terms of medians of f1-scores as well as their distributions. 

While this gives a good idea of how well a classifier would actually perform if trained on our dataset, we could not use more thorough statistical comparisons between classifiers. This is because a method such as analysis of variance for comparing means requires the observations to be independent of each other. The classifiers used to generate the f1-scores across the 20-folds have overlapping training observations for most of the runs, which results in the lack of independence of the generated f1-scores.

\subsection{Baseline Models}

For this research, we set up our baseline using a few traditional machine-learning models. These models are Support Vector Machine (SVM), Support Vector Machine using Stochastic Gradient Descent (SGD), Multinomial Na\"{i}ve Bayes (MNB), Logistic Regression (LR), Random Forest (RF), Gradient Boosting (GB), and AdaBoost (AB).

\subsubsection{Input Embedding}

The input to our baseline models was first processed by the TF-IDF vectorizer in scikit-learn \cite{pedregosa2011scikit}. TF-IDF vectorization is a common way to convert raw text and documents into embeddings that are suitable for machine-learning models. The vectorizer generates a document-vocabulary matrix for each of the documents (in our case, review comments).  Then, using inverse document frequency, it normalizes ("lowers") the weight of the words by checking how often a word appears in other documents (comments, in this case). This helps lessen the impact of frequent yet unimportant words for our classification task, so that common words like "the" that convey little semantic meaning do not affect the classification of a comment.

\subsubsection{Support Vector Machine}
Support vector machines are commonly used for classification in machine learning. A SVM establishes a decision boundary as well as a positive plane and a negative plane between classes.  Anything in the positive plane is considered to have the characteristic under study (in our case, it is considered to state a problem with the work that is being reviewed).

The goal of the training process is to maximize the distance between the positive and negative planes, so that data (in our case, review comments) belonging to the two classes are best separated. 

\begin{figure}[htbp]
  \centering
  \includegraphics[width=0.7\linewidth]{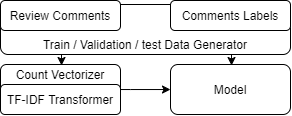}
  \caption{Data pipeline for machine learning model}
  \label{fig:svm_model}
\end{figure}

We created a data pipeline that fed the sentences from a review comment (an average of 2.2 sentences per comment) into a count vectorizer, then transformed values coming out of it with a TF-IDF transformer, before feeding it into the SVM classifier together with the label corresponding to that sentence. This pipeline is shown in Figure \ref{fig:svm_model}. Statistical features for each review comment represented in TF-IDF-normalized vectors are put into the vector space for all comments, then the model learns a hyper-plane (support vector) to best divide them into two categories: comments containing problem statements, and comments without problem statements. The SVM kept learning from inputs in multiple rounds until it could no longer improve its accuracy.

To improve model accuracy, we tuned the inverse regularization parameter \textit{C}, as well as vocabulary length in the count vectorizer. Through hyper-parameter grid search we found that the SVM model reaches its best accuracy with \textit{C} = 1 and bi-gram as vocabulary. 

\subsubsection{Support Vector Machine with Stochastic Gradient Descent}
Current developments in machine-learning techniques allowed us to modify some of the existing machine-learning models to enhance their accuracy. Stochastic Gradient Descent (SGD) was developed early on and popularly adopted to optimize neural-network models \cite{lecun2012efficient}, while applying SGD on linear classifiers is not unheard of. \cite{ruder2016overview} We compared the performance of the SVM model with and without SGD.

In this model, we applied a combination of L1 and L2 regularization to the loss function, with the hope of correcting over-fitting problems. The ratio between L1 and L2 regularization is a hyper-parameter that we tuned with results from the validation set.

The implementation of this model is very similar to the baseline SVM model, as shown in Figure \ref{fig:svm_model}, except for the classifier that utilizes SGD as an optimizer. Hyper-parameters tuned for this model include the regularization modifier $\alpha$, the ratio between the two normalization functions, and vocabulary length in the count vectorizer. The model is  optimal when $\alpha$ = 0.001, with 0.15 regularization, and bi-gram as vocabulary.

\subsubsection{Multinomial Naive Bayes}

A na\"{i}ve Bayes model assumes that each of the features it uses for classification are independent of one another. 
\[p(f1,...,fn|c)=\prod_{i=1}^{n}p(fi|c)\] 
A multinomial na\"{i}ve Bayes Classifier is a special instance of a NB classifier that follows a multinomial distribution for each feature $p(fi|c)$.
In order to determine whether a review comment identifies a problem, the model examines the TF-IDF normalized word-count vectors for that comment, using the conditional probability of each of these features/vectors, and makes a judgment, based on conditional probabilities learned from the training set.

To get the optimal model accuracy, we tuned smoothing parameter $\alpha$ as well as vocabulary length in the count vectorizer.  The model achieved its best accuracy with an $\alpha$ of 1 and bi-gram as vocabulary.  

\subsubsection{Logistic Regression}

The logistic-regression (LR) classifier uses a regression equation to produce discrete binary outputs. Very similarly to linear regression, it learns the coefficients of each input feature through training; however it uses a logistic function instead of linear activation to determine the class to which an input belongs.

As can be seen in Figure \ref{fig:svm_model}, in our case, the LR classifier learns coefficients of each word through comments in the given training set, and then in the testing stage, uses those coefficients together with logistic activation to determine whether a comment contains a problem statement. During training, a few hyper-parameters are tuned through grid search: the inverse regularization parameter \textit{C}, the solver/optimizer, as well as vocabulary length in the count vectorizer. We found through hyper-parameter grid search  that the LR model achieves its best accuracy with a \textit{C} = 10, lib-linear solver and bi-gram as vocabulary. 

\subsubsection{Random Forest}

The Random Forest (RF) classifier is an ensemble method that fits multiple decision trees and uses averaging to improve the accuracy of predictions as well as to avoid over-fitting. 

To get the best performance out of our RF model shown in Figure \ref{fig:svm_model}, we tuned a few hyper-parameters associated with it: the metric to track quality of each split, the number of trees to be used, max depths of these decision trees, as well as vocabulary length in the count vectorizer. We found through hyper-parameter grid-search that the RF model achieves its best accuracy with Gini Impurity measurement to determine split, and 300 trees with maximum of depths of 100, and uni-gram as vocabulary.

\subsubsection{Gradient Boosting}

Gradient boosting (GB) is an ensemble machine-learning algorithm that utilizes a number of weak models, such as small decision trees. Unlike random-forest algorithms, which use fully grown decision trees, decision trees in GB models use are normally in limited numbers and depth. In training, these small decision trees are fitted in a negative gradient direction in order to reduce the loss calculated from the cost function.

We have tuned the learning rate and vocabulary length in the count vectorizer in order to achieve better performance. Results showed that the model achieved its best accuracy with a learning rate of 0.3, 150 estimators, and bi-gram as vocabulary.

\subsubsection{AdaBoost}

AdaBoost, or adaptive boosting, is a meta-algorithm that alters weights of entries for base models. When an entry is misclassified, the algorithm increases the weight of that entry and subsequently decreases the weights of entries that have been correctly classified.  The algorithm terminates upon meeting the confidence threshold. Through doing this, the booster identifies the features that have greater impact on the results, and improves prediction accuracy.

As can be seen in Figure \ref{fig:svm_model}, in the case of text classification, comments are presented to the classifier in the form of normalized word vectors. The base model trains on the input with the AdaBooster. When a comment is misclassified, the booster increases the weight of that comment, while decreasing weights of comments that are classified correctly. Thus, the base model uses this new information to better train itself.

To obtain the optimal accuracy from our AdaBoost model, we tuned the learning rate of the model, the number of trees to be used, as well as vocabulary length in the count vectorizer. We found through grid search that AdaBoost model achieves its best accuracy with 0.8 as learning rate, 170 trees and bi-gram as vocabulary.

\subsection{Neural Network Models}

Our other experiments use neural networks, and Keras \cite{chollet2015keras} was the framework of choice for implementation. Compared with our baseline models, the input of each model is generated in two different ways: through a GloVe embedding and BERT embedding.

\subsubsection{Input Embedding}

Global Vectors for Word Representation, or GloVe embedding \cite{pennington2014glove}, is an embedding model that converts words into multidimensional vectors based on their meaning. Its function is similar to Word2Vec, which transforms words to embeddings in a limited vector space, though the underlying principle is different. Word2Vec utilizes a continuous bag of words (CBOW) to reconstruct a word using context in a window surrounding it, essentially as an auto-encoder. The bottleneck layer of this auto-encoder is used to embed given words in Word2Vec, while GloVe utilizes global counts of words, then factorizes this data into lower dimensions and uses that to embed the given words.

Bidirectional Encoder Representations from Transformers (BERT) is a multi-layer bidirectional transformer encoder \cite{devlin2018bert} developed by Google. The original BERT paper describes how the model is used to perform two tasks: predicting masked words in sentences, and classifying whether two sentences are next to each other in a document. It utilizes multiple layers of bidirectional transformers and classification layers to achieve its goal. To use it as a sentence embedding layer, one could extract a number of the transformer layers of a trained model with frozen weights, then input a sentence and use its output as a sentence embedding.

The BERT network we used in our experiment is published by Google and is pre-trained on Wikipedia and BooksCorpus data. We used the open-source project "Bert-as-service" to create sentence embeddings.  Specifically, we limited the maximum sentence length to 25 words, and extracted embeddings with outputs from the second-last layer in the pretrained network. The Bert-Base-Uncased model \cite{devlin2018bert} has 12 attention layers, and 768 neurons in each layer with 12 attention heads. Using this network has given us 768 dimensions as  sentence embeddings. We also used a version with word level embeddings.

In the next subsection, we are going to talk about how we utilized these embeddings in our neural network models to perform the classification task.

\subsubsection{Multilayer Perceptron}

A multilayer perceptron (MLP) model \cite{hastie2005elements} is a typical artificial neural network. It utilizes multiple layers of neurons, and uses back-propagation for training. Errors calculated by a loss function are propagated back through the layers using the chain rule of gradient descent derivation. MLP is often referred to as a "vanilla" type of neural network, for its simple mathematical background and structure.

\begin{figure}[htbp]
  \centering
  \includegraphics[width=0.7\linewidth]{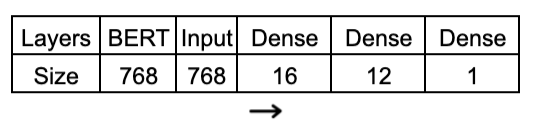}
  \caption{Data pipeline for Multilayer Perceptron with BERT model}
  \label{fig:mlp_model}
\end{figure}

Using a BERT embedding, we converted each sentence into a 768-dimensional vector, then fed it into a shallow MLP with three layers, as shown in Figure \ref{fig:mlp_model}

\subsubsection{Convolutional Neural Network}

A convolutional neural network (CNN) utilizes convolution kernels that pool data with a defined window size together on given dimensions to generate summaries from input data \cite{goodfellow2016deep}. While CNNs are widely used for computer vision, and occasionally in natural language processing, they can also be used in extracting information, such as {\it n}-grams (in the time dimension) and dimension summarizing (in the feature dimension).

\begin{figure}[htbp]
  \centering
  \includegraphics[width=1\linewidth]{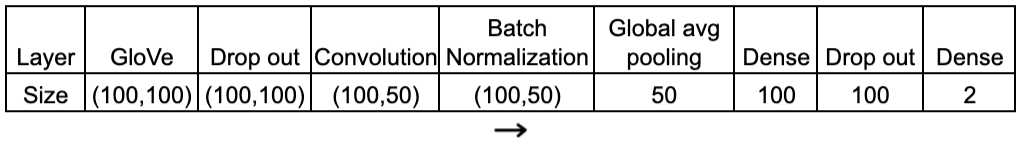}
  \caption{Data pipeline for CNN Model}
  \label{fig:cnn}
\end{figure}

When dealing with comment classification, this model (Figure \ref{fig:cnn}) uses convolutions on the feature dimension to reduce the complexity of each word vector, different dropout percentages, and pooling methods. Varying activation functions and optimizers have been attempted to achieve the best result. We found the best result came with 20\% and 40\% dropout rate in the two dropout layers, ReLU activation function for the hidden layer, softmax for output layer, global average pooling for pooling layer, and ADAM as optimizer.

\subsubsection{Recurrent Neural Network}

Recurrent neural networks (RNNs) are neural networks that take time-sequence information into consideration.  For each time-step, the network takes the inputs and updates its internal memory cells with new information. Different RNN models implement memory updates differently.  For example, long short-term memory (LSTM) networks not only remember inputs, but "forget" information that is not important as well. 

Generally RNNs use a few gates for internal memory manipulation. A "vanilla" RNN uses a memory gate to update its internal memory state for each input, while it has a problem with gradient vanishing going through excessive amount of time steps. A LSTM network applies more gates, including a "forget gate," an "input gate," and an "output gate" to solve this problem. By adding these additional gates, the network can choose to "forget" what's not important to the result, and only keep meaningful information in its internal memory state.  A gated recurrent unit (GRU) network does this with only two gates, a "reset gate" and an "update gate."

When we pass an embedded sentence to the network, each word is seen as an item emerging in one time step, and the sequence of words in a sentence becomes a sequence of vectors transitioning along with time steps. The neural network learns from the transition what information is important to keep versus what is not, then applies the same judgment when a new sentence is given to it for classification.

\begin{figure}[htbp]
  \centering
  \includegraphics[width=0.8\linewidth]{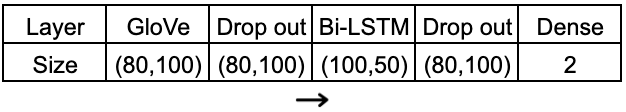}
  \caption{Data pipeline for Bi-LSTM Model}
  \label{fig:lstm}
\end{figure}

The network structure of our implementation can be found in Figure \ref{fig:lstm}. In order to receive optimal results, we used a bidirectional LSTM layer. Our experiments showed that adding the extra direction did benefit the result, and the model's best performance is recorded under the following settings: 40\% and 30\% dropout rate on the two dropout layers, ReLU activation function for the hidden layer, softmax for output layer, with ADAM as optimizer.

Here we also implemented a GRU network and a bidirectional GRU network in parallel, with similar structure shown in Figure \ref{fig:lstm}, results could be found in Section \ref{section_results}

\subsubsection{Hierarchical Attention Network}\label{HAN Section}

Hierarchical attention networks (HANs) are neural networks that take into consideration the document structure and sentence structure \cite{yang2016hierarchical}. A document normally consists of a number of sentences, and a sentence is formed by a number of words. Not all sentences in a document are important to the classification of a document, and similarly, not all words are important for sentence-level classification. HANs utilize this information through attention layers that capture words and sentences that are important towards the classification. HANs have reached accuracies in the high 60\%s and lower 70\%s on the Yelp 2013-2015 datasets.

In classifying comments, a HAN can capture information with greater impact on the results. For example in sample comment "The writeup does not include a Test Plan section," the words "does not include" contributes a lot more to implying there is a problem stated in this comment than other parts of the comment do.

\begin{figure}[htbp]
  \centering
  \includegraphics[width=0.9\linewidth]{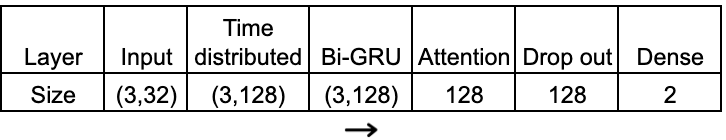}
  \caption{Data pipeline for HAN Model}
  \label{fig:han}
\end{figure}

The HAN model (Figure \ref{fig:han}) reached its best performance on the following settings: 40\% and 30\% dropout rate on the two dropout layers, ReLU activation function for the hidden layer, sigmoid for output layer, with ADAM as optimizer.

\subsubsection{Convolutional Neural Network with Long Short Term Memory}

Previous models showed that each type of the neural network or neural network layer could be efficient on specific tasks, for example CNN for dimension reduction and HAN for extracting words that are more important to the result. In this subsection we are going to combine some of the models and explore the benefits of mixing different types of neural networks.

\begin{figure}[htbp]
  \centering
  \includegraphics[width=1\linewidth]{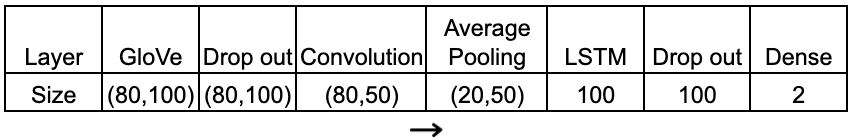}
  \caption{Data pipeline for CNN + LSTM Model}
  \label{fig:cnn_lstm}
\end{figure}

In Figure \ref{fig:cnn_lstm}, a model with CNN and LSTM layers is implemented in the hope of securing benefits from both models. With CNN as a dimension reducer, the LSTM layer might be able to find useful information from the aggregated features. Results are found in Section \ref{section_results}



\begin{figure}[htbp]
  \centering
  \includegraphics[width=1.05\linewidth]{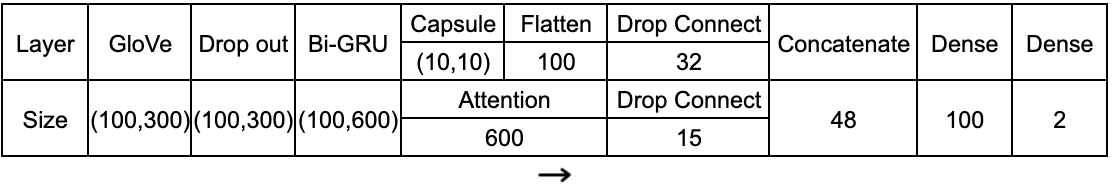}
  \caption{Data pipeline for Bi-GRU + Attention + Capsule Model}
  \label{fig:bigru_cap_att}
\end{figure}

Another attempt shown in Figure \ref{fig:bigru_cap_att} tests whether having a CNN to reduce dimensions is necessary, by removing it while boosting the performance of recurrent layer by putting it in a bidirectional wrapper.



\section{Experimental Results} \label{section_results}
\subsection{Baseline Models}
\begin{figure}[htbp]
  \centering
  \includegraphics[width=\linewidth]{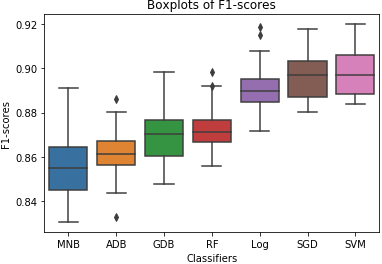}
  \caption{Baseline Models F-1 Scores}
  \label{fig:baseline_box}
\end{figure}

Figure \ref{fig:baseline_box} displays a boxplot of the 20 f1-scores obtained using the traditional machine-learning classifiers from the 20-fold cross validation. F1-score is the harmonic mean of precision and recall. In the case of datasets with balanced classes, the f1-score also represents the classifier accuracy on the test set. The lowest-performing classical machine learning classifiers, multinomial na\"{i}ve Bayes and AdaBoost, achieved similar accuracy, with respective sample median f1-scores of 0.855 and 0.861. The gradient boosting and random-forest classifiers achieved sample median f1-scores of 0.870 and 0.871. The highest performing classifiers included logistic regression, stochastic gradient descent, and support vector machines. They achieved sample median f1-scores of 0.890, 0.897, and 0.897 respectively.

The sample standard deviation of f1-scores for each traditional classifier tended to range about a 1\% difference from the sample mean. Along with the sample minimum and maximum, the sample standard deviation indicates that there was not a large variance in classifier accuracy on the various folds of the cross validation. While some outliers were present, as indicated in the boxplot from Figure \ref{fig:baseline_box}, they did not skew the sample mean far enough from the sample median score to give a different impression of the classifier's performance. These results show that baseline classifiers can classify review comments as detecting problems with an accuracy range of approximately 85\% to 89\%.

To gain insight into the phrases that contributed towards determining a suggestion, we extract coefficient weights of some features from two of the models. Table \ref{tab:log_coefficients} displays a list of the logistic regression model's top 10 positive and negative features in determining if a comment has mentioned a problem in the author's work. The features that increase the likelihood that a comment will mention a problem (positive coefficients) include some phrases that may constitute a suggestion by the reviewer. For instance, phrases such as "could", "should", "could have", and "more" indicate that the reviewer is likely giving advice to the author about improving the work, thus noting a problem by implication. Features with negative coefficients include phrases that likely demonstrate positive sentiment, such as "yes", "good", "well", and "great".

 \begin{table}[h!]
 \caption{Logistic Regression Coefficients}
  \centering
  \begin{adjustbox}{max width=\textwidth}
  \begin{tabular}{*{4}{|c}|}
  \hline
  Coefficient &  Value & Coefficient & Value\\
  \hline
  \hline
  yes & -8.0233 & not & 10.5227\\
  good & -3.9472 & but & 8.8498\\
  and & -3.1690 & however & 7.8254\\
  they have & -3.1193 & more & 6.2155\\
  well & -3.0567 & could & 5.6703\\
  yes the & -2.9953 & should & 5.3498\\
  all the & -2.7422 & would & 5.0391\\
  clearly & -2.6269 & no & 5.0183\\
  project & -2.5331 & missing & 4.9864\\
  passed & -2.4645 & some & 4.9160\\
  \hline
\end{tabular}
\end{adjustbox}
  \label{tab:log_coefficients}
\end{table}

Table \ref{tab:sgd_coefficients} displays the stochastic gradient descent model's top 10 positive and negative features in determining if a comment mentioned a problem in the author's work. The coefficient values are lower than  those of the logistic regression model. However, they comprise similar positive and negative features.

\begin{table}[h!]
 \caption{Stochastic Gradient Descent Coefficients}
  \centering
  \begin{adjustbox}{max width=\textwidth}
  \begin{tabular}{*{4}{|c}|}
  \hline
  Coefficient &  Value & Coefficient & Value\\
  \hline
  \hline
  yes & -4.1029 & however & 6.5277\\
  conflicts & -2.0396 & not & 6.4184\\
  good & -2.0083 & but & 5.5175\\
  apply & -1.7788 & should & 3.9721\\
  complicated & -1.7785 & could & 3.9198\\
  since & -1.6178 & would & 3.8352\\
  sense & -1.6139 & more & 3.6346\\
  required & -1.5925 & missing & 3.5942\\
  passed & -1.5757 & no & 3.4112\\
  project & -1.5637 & except & 2.9776\\
  \hline
\end{tabular}
\end{adjustbox}
  \label{tab:sgd_coefficients}
\end{table}

\begin{figure}[htbp]
  \centering
  \includegraphics[width=\linewidth]{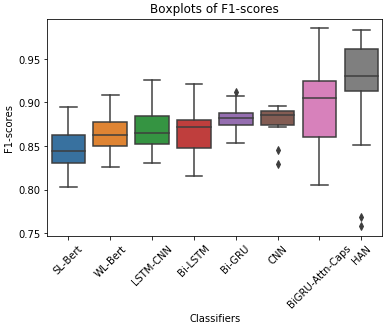}
\begin{center}
\begin{tabular}{ r c r c}
\footnotesize SL-Bert & \footnotesize 0.844 & \,\,\footnotesize Bi-GRU & \footnotesize 0.882 \\
\footnotesize WL-Bert & \footnotesize 0.862 & \,\,\footnotesize CNN & \footnotesize 0.886 \\
\footnotesize LSTM-CNN & \footnotesize 0.865 & \,\,\footnotesize BiGRU-Attn-Caps & \footnotesize 0.905 \\
\footnotesize Bi-LSTM & \footnotesize 0.872 & \,\,\footnotesize HAN & \footnotesize 0.931 \\
\end{tabular}
\end{center}
  \caption{Neural Network Models Median F-1 Scores}
  \label{fig:nn_box}
\end{figure}

\subsection{Neural Network Models}
Figure \ref{fig:nn_box} shows the results for the neural-network architectures. Combining the results from different permutations of input data through 20-fold cross validation gives us a better view of how the models performed than using the f1 score from just one experiment. Therefore, we use the median f1-scores from the result distributions to gauge classifier performance. The sample standard deviation of f1-scores for the neural network classifiers tended to range around a 2\% difference from the sample mean, with the exception of the BiGRU-Attn-Caps and HAN models with around 4\% and 5\% respectively. The larger variance in these models than in those with the traditional classifiers indicates that we will be less sure of the actual performance that the neural network classifiers would achieve in a real setting than we would with the traditional classifiers. The CNN  classifier produced several notable outliers, while the HAn classifier produced two extreme outliers. Compared with the models that use BERT, the models using GloVe embeddings appear to have had less trouble finding the feature relationships that determine whether a comment mentions a problem.

As can be seen in Figure \ref{fig:nn_box}, around half of the neural networks are performing slightly worse than the baseline models, with the others beings being around or notably greater than the baseline. The best two performing neural networks, being the BiGRU-Attn-Caps and HAN models were also better than the traditional classifiers in detecting the presence of a problem in a comment.

The HAN and BiGRU-Attn-Caps models that used GloVe embeddings achieved the best performance among all the models. The CNN model that used GloVe embeddings achieved the next best performance with a sample median f1-score of 0.886. The Bidirectional GRU had a very close sample median f1-score of 0.882, followed by the Bidirectional LSTM model with 0.872, then the LSTM CNN model at 0.865. The lowest-scoring models were the ones with word-level (WL-Bert) and sentence-level (SL-Bert) BERT embeddings with sample median f1-scores of 0.862 and 0.844 respectively.

\section{Conclusions and Future Work}

We have marshalled a multitude of classifiers that can parse student peer-review comments for the detecting the mention of a problem. The HAN and BiGRU-Attn-Caps models performed the best among the neural network classifiers on this dataset, while the best traditional classifiers were the support vector machine and stochastic gradient descent models. The logistic regression classifier, CNN and Bidirectional GRU models, were the next best. This was followed by most of the remaining classical machine learning classifiers using TF-IDF features, with the remaining array of neural network models being close to or approximately equal in score. The least effective classical models were the AdaBoost and multinomial na\"{i}ve Bayes classifiers. The least effective neural network models were the two that used the sentence and word level embeddings.

Future research will go in two directions. First, we will further explore the potential of our models, both those using BERT embeddings and GloVe embeddings, in other text-classification tasks related to peer reviews. Such tasks include identifying arguments made by the reviewer. We also plan to use these models to give real-time feedback to the reviewer before the review is submitted. This feedback will compare the about-to-be-submitted review with other reviews in the system, with respect to the detection of problems, arguments, and so forth. This will allow the reviewer to improve the review before submission. This also enables automatic grading of peer review feedback in settings where such functionality would be desirable. The final step will be to measure the efficacy of reviews pursuant to this feedback with reviews that do not have the benefit of such feedback. Our long-term goal is to deploy these models in a classroom setting to measurably improve the quality of reviewing, and hence of student learning.

\bibliographystyle{unsrt}
\bibliography{problem_detection_complete}

\end{document}